\documentclass[12pt]{iopart} 
\usepackage{iopams}
\usepackage{cite}

\usepackage[utf8]{inputenc}
\usepackage{graphicx}
\usepackage{xcolor}

\usepackage{soul}  

\usepackage{bm}
\renewcommand{\vec}{\bm}
\begin{document}

\title{Inverse patchy colloids with small patches: fluid structure and dynamical slowing down}
\author{ Silvano Ferrari$^1$, Emanuela Bianchi$^1$, Yura V. Kalyuzhnyi$^2$, Gerhard Kahl$^{1,3}$}
\address{$\phantom{1}^1$ Institut f{\"u}r Theoretische Physik, Technische Universit{\"a}t Wien, Wiedner Hauptstra{\ss}e 8-10, A-1040 Wien, Austria}
\address{$\phantom{1}^2$ Institute for Condensed Matter Physics, Svientsitskoho 1, 79011 Lviv, Ukraine}
\address{$\phantom{1}^3$ Center for Computational Materials Science (CMS), Technische Universit{\"a}t Wien, Wiedner Hauptstra{\ss}e 8-10, A-1040 Wien, Austria}
\ead{silvano.ferrari@tuwien.ac.at}

Date: \today

\begin{abstract}
Inverse Patchy Colloids (IPCs) differ from conventional patchy particles because their patches repel (rather than attract) each other and attract (rather than repel) the part of the colloidal surface that is free of patches.
These particular features occur, .e.g., in heterogeneously charged colloidal systems.
Here we consider overall neutral IPCs carrying two, relatively small, polar patches. Previous studies of the same model under planar confinement have evidenced the formation of branched, disordered aggregates composed of ring-like structures.
We investigate here the bulk behavior of the system via molecular dynamics simulations, focusing on both the structure and the dynamics of the fluid phase in a wide region of the phase diagram.
Additionally, the simulation results for the static observables are compared to the Associative Percus Yevick solution of an integral equation approach based on the multi-density Ornstein-Zernike theory.
A good agreement between theoretical and numerical quantities is observed even in the region of the phase diagram where the slowing down of the dynamics occurs.
\end{abstract}

\submitto{\JPCM}

\maketitle

\section{Introduction}\label{sec:introduction}

In the last ten years, patchy colloidal
particles~\cite{ilona:review,bianchi:review} have emerged as one of
the most promising solutions to the problem of designing new materials
via a bottom-up strategy.  Due to the
specificity of their bonding patterns, patchy particles have indeed
proven to be perfect candidates as nano- and micro-scale building
blocks for materials with desired symmetries
and physical properties~\cite{glotzer:review}. The interest
in these systems is triggered by the wide
range of potential applications, e.g. in photonics, biomimetic
materials synthesis or electronics, and relies on the impressive
advances in the production of colloidal particles with tunable
chemical/physical surface designs~\cite{ilona:review}.  Even though
experimental studies on assembly processes of patchy particles are
so far rather rare~\cite{velegol2009,velev2010,granick2011,pine2012}, there is no
doubt about the great potentialities of these colloids as
self-assembling units of completely new macroscopic structures. In
fact, numerical and theoretical studies have opened the way to a
surprising multitude of both ordered and disordered
phases~\cite{bianchi:review} that significantly extends the many
possibilities already offered by isotropically interacting colloidal
particles.
 
A further direction of investigation in the field of patchy particles has been recently put forward by the introduction of patchy systems with charged surface regions; these systems are also referred to as Inverse Patchy Colloids (IPCs)~\cite{bianchi:ipcfirst}. While conventional patchy particles are built by adding
attractive patches on the top of otherwise repulsive spherical
particles, {\it inverse} patchy particles
have patches that repel each other, while they attract
the rest of the colloidal surface that is free of patches.  IPC model
systems are thus suitable to study self-assembling units with a
non-homogeneous surface charge distribution.  Within this broad class of systems, we focus here on
heterogeneously charged colloids with two charged polar patches and an
oppositely charged equatorial belt.  The analytical description of the
underlying microscopic system has been provided in
Ref.~\cite{bianchi:ipcfirst}; it features a
complex expression in terms of Bessel functions and Legendre
polynomials for the pair potential between two of such IPCs.  The
related coarse-grained description, also presented in
Ref.~\cite{bianchi:ipcfirst}, allows to deal with a much simpler model
that maintains the key physical features of the microscopic systems.
The model was originally developed to describe the interaction between
complex units emerging from the adsorption of charged polyelectrolyte
stars onto the surface of oppositely charged colloids, but it works
equally well to describe heterogeneously charged particles
synthesized in the lab~\cite{peter}.  The only prerequisite of the
applicability of our description -- analytical as well
as coarse-grained -- is that the center
of charge of a given surface region does not lie outside the
colloidal particle.

The equilibrium behavior of IPCs is dominated by a non-trivial
interplay between attractive and repulsive directional interactions.
A very recent example of exotic structures emerging in IPC systems in
the bulk has been reported in Ref.~\cite{ismene}.  There, slightly
overall charged IPCs with two, relatively extended and long-ranged,
polar patches were selected and their phase diagram was evaluated by
means of Monte Carlo simulations and free energy calculations~\cite{evagunther,evamethod}.
In addition to the fluid phase and to the fcc (ordered and plastic) solids, the resulting
phase diagram features a wide region in temperature and density where
a crystal formed by parallel monolayers is stable; the stability of this laminar phase was also investigated on changing the interaction parameters~\cite{evanew}. 
The effect of the interaction range and of the patch extension was also studied in the bulk fluid phase: the structural and thermodynamic properties of a selection of overall neutral IPC systems have been studied by means of Monte Carlo simulations and theoretical approaches.
In particular, the multi-density Ornstein-Zernike integral equation, supplemented with the Associative Percus Yevick (APY) approximated closure~\cite{wertheim1986c,wertheim1986d,wertheim1987,kalyuzhnyi1991}, has been extended to describe IPCs with an arbitrary number of patches.
From the comparison between the simulation data and other theoretical approaches,
such as the Reference Hypernetted Chain
description~\cite{Lado,Giacometti} and the Barker-Henderson
thermodynamic perturbation theory~\cite{bhoriginal,Gogelain}, it
emerged that the APY approach is a better candidate to describe
the IPC class of systems, performing a higher accuracy in the
results and a wider convergence ability when lowering the system temperature.

In the present paper, we focus on overall neutral IPCs with relatively small patches and a short interaction range. Under planar confinement conditions and at intermediate densities, the selected system did not form ordered structures, but rather disordered, branched aggregates composed of particle rings~\cite{bianchi:2d2013}.
In contrast, IPCs with the same interaction range and particle charge, but bigger patch size tend to form planar colloidal monolayers with a broad variety of translational and orientational ordering~\cite{bianchi:2d2013,bianchi:2d2014}.

In this contribution, we consider the relatively small patch case and we focus on its bulk behavior in the fluid phase, while the bulk counterpart of the other models investigated under planar confinement~\cite{bianchi:2d2013,bianchi:2d2014} will be addressed in future publications.
We use Molecular Dynamics (MD) simulations to study a selected IPC system, focusing on the structural and dynamical properties of the fluid phase in a wide range of temperatures and densities, including the region where the dynamical slowing down occurs. Additionally, we compare the numerical results for the static properties at different state points with the APY theoretical description.

Our goal is twofold: on one side we want to check how far the APY theory can be extended to the high-density and/or low-temperature regions; on the other side we want to study the behavior of the system in the region where the dynamical slowing down occurs.

The paper is organized as follows.
In section~\ref{sec:model} we introduce the IPC model.
In section~\ref{sec:methods} we describe the investigation methods, with particular attention to the way the interaction potential was implemented in the MD simulation code.
In section~\ref{sec:results} we describe our results, namely the static properties are discussed in section~\ref{sec:static}, while the dynamic ones are reported in section~\ref{sec:dynamics}.
Our final considerations are presented in section~\ref{sec:conclusions}.

\section{Model and methods}\label{sec:model_methods}

\subsection{Model}\label{sec:model}

\begin{figure} 
\begin{center}
\includegraphics[width = 1.0\textwidth]{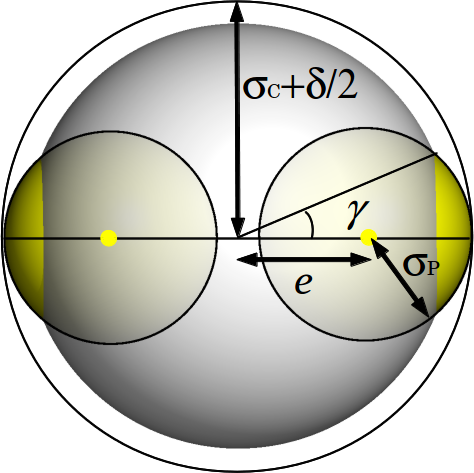}
\caption{
Schematic representation of the IPC model with two patches.
The dark gray sphere is the hard sphere (HS) colloid with radius $\sigma_{\rm c}$.
The yellow dots inside the particle at distance $e$ from the particle center are the interaction sites, while the yellow spheres, partially located inside the HS core, represent the interaction zone of the patch  with radius $\sigma_{\rm p}$; only the cap of the yellow sphere that emerges from the HS core can interact with other particles, thus defining the patch angular semiamplitude $\gamma$.
The  external black circle of radius $\sigma_{\rm c}+\delta/2$ represents the colloid interaction sphere.
Geometric constraints impose that the interaction range is given by $\delta/2=e+\sigma_{\rm p} - \sigma_{\rm c}$.
\label{fig:ipcmodel}}
\end{center}
\end{figure}

Our two-patch IPCs are modeled as hard spheres of radius $\sigma_{\rm c}$ carrying three interaction sites: one corresponding to the colloid center and two corresponding to the patch centers; the latter are placed at a distance $e$ from the first, in opposite directions (see Figure~\ref{fig:ipcmodel}).
The bare colloid interaction sphere has radius $\sigma_{\rm c}+\delta/2$, where $\delta$ is the resulting interaction range, while both the patch interaction spheres have radius $\sigma_{\rm p}$; $2\sigma_{\rm c}$ is assumed to be the unit of length.
By construction the following relations hold~\cite{bianchi:ipcfirst}
\begin{equation}
  \sigma_{\rm c} + \delta/2 = e + \sigma_{\rm p} \qquad {\rm and} \qquad
   \cos \gamma = \frac{\sigma_{\rm c}^2 +e^2 -\sigma_{\rm p}^2}{2e\sigma_{\rm c}},
\end{equation}
where $\gamma$ is half of the patch opening angle.
By virtue of these constraints, the model is characterized only by two geometrical parameters: the interaction range $\delta$ and the patch extension $\gamma$.

Beyond the hard core repulsion, the pair potential between two IPCs at distance $r$ is given by~\cite{bianchi:ipcfirst}
\begin{equation}
U  = 
\left\{
\begin{array}{rl}
\frac{3\sigma_{\rm c}^3}{4\pi}\sum_{ij} u_{ij}w_{ij} &{\hspace{1em}\rm if\hspace{1em}} 2\sigma_{\rm c}<r<2\sigma_{\rm c}+\delta \\
0                          &{\hspace{1em}\rm if\hspace{1em}} r \ge 2\sigma_{\rm c}+\delta,
\end{array}
\right.
\end{equation}
where $i$ ($j$) specifies one of the three interaction sites of the first (second) IPC, $w_{ij}$ is the overlap volume of the corresponding interaction spheres, and $u_{ij}$ is the energy strength of the $ij$ interaction.
We note that, while the $u_{ij}$ are constants, the $w_{ij}$ -- as well as the potential $U$ -- depend on both the inter-particle distance and the relative orientation of the two IPCs; we drop the arguments for brevity.

For two identical patches, the set of the $u_{ij}$ reduces to three independent energy constants: $u_{\rm cc}$, $u_{\rm pp}$ and $u_{\rm cp}$.
The coarse-grained model parameters $\delta, \gamma$ and $\bar{u}=(u_{\rm cc}$, $u_{\rm pp}$, $u_{\rm cp})$ for a specific microscopic system can be fixed by taking advantage of the Debye-H{\"u}ckel analytical description of the underlying model~\cite{bianchi:ipcfirst}.

In the present contribution, we consider IPCs under relatively high screening conditions, namely we fix $\kappa \sigma_{\rm c}=5$, where $\kappa^{-1}$ is the Debye screening length, and $\kappa\delta=2$; the corresponding interaction range is $\delta=0.4\sigma_{\rm c}$.
The polar patches are chosen to be relatively small, namely $\gamma \approx 30^\circ$, corresponding to the choice $e=0.64\sigma_{\rm c}$ and $\sigma_{\rm p}=0.56\sigma_{\rm c}$.
The array $\bar{u}$ results from considering overall neutral colloids of diameter 60nm, dispersed in water at room temperature.
The energy minimum, obtained when two particles at contact are in a T-shape configuration, sets the unit of energy $\epsilon_m$. These features lead to $\bar{u}=(0.8631, 241.8,-26.62)$ in units of $\epsilon_m$. In the following we use reduced units for the temperature, $T^*=k_{\rm B}T/\epsilon_{m}$, the energy per particle, $u^*=U/(N\epsilon_m)$, the chemical potential $\mu^*=\mu/\epsilon_m$ and the density, $\rho^*=\rho(2\sigma_{\rm c})^3$.

\subsection{Methods}\label{sec:methods}

The bulk fluid phase of our IPC system was investigated with MD simulations in the NVE ensemble using the velocity Verlet integration scheme~\cite{frenkelsmit} and the RATTLE algorithm for geometric constraints~\cite{RATTLE}.
Since the conventional RATTLE procedure is singular for linear particles~\cite{ciccotti}, we describe the patch-center-patch complex reducing the number of degrees of freedom according to the scheme introduced in Ref.~\cite{ciccotti}.
We proceed as follows: the two patches are treated as two particles separated by a fixed distance and move according to effective forces that include both the inertia of the central particle and the forces acting on it.
As a consequence of these effective forces, the central particle automatically satisfies the constraint of being in the middle point of the line joining the two patches and its trajectory is automatically provided by the knowledge of the patch trajectories.
This approach reduces the number of equations of motion from $18N$ to $12N$ for a  system of $N$ IPCs.

In order to integrate the equations of motion, we approximated the hard core repulsion via a continuous harshly repulsive soft sphere interaction
\begin{equation}
  U(r) = A \left [ \left(\frac{2\sigma_{\rm c}}{r}\right)^{2k}-2\left(\frac{2\sigma_{\rm c}}{r^{k}}\right)^{k}+1 \right ] \Theta(2\sigma_{\rm c}-r)
\end{equation}
with $k=15$ and $A=500$.
We checked the consistency between the continuous and the hard core~\cite{bianchi:ipcfirst} versions of the model by comparing energies, pair distribution functions and static structure factors at several state points with those obtained via Monte Carlo simulations of the original model~\cite{bianchi:2d2013}. No significative differences were observed.

Since  in the microcanonical ensemble the temperature fluctuates, our temperature values are calculated as time averages over the kinetic temperature.

The algorithm requires to specify a mass for each interaction center.
We chose to assign the same mass, $m_0$, to the central particle as well as to each of the two patches.
This choice fixes the total mass of an IPC to $3m_0$ and its inertia tensor is thus $I={\rm diag}(2m_0e^2,2m_0e^2,0)$.
Defining $m_0$ as the mass unit also fixes the time unit, i.e. $t_0 = \sqrt{ m_0(2\sigma_{\rm c})^2/\epsilon_m }$.

The timestep of the simulation was chosen to be $2.5 \times 10^{-3}$ in units of $t_0$; this choice ensured that the total energy was conserved within $10^{-2}$ \% of its mean value.

Most results shown in this contribution were obtained with a system of 2048 particles; on few occasions ensembles with 4000 were considered.
To check for possible size effects, simulations for $\rho=0.40$ state points were carried out for both ensemble sizes. All observables agreed within numerical accuracy: for instance the internal energy showed differences only in the third digit.
The systems were equilibrated for over $10^5 t_0$, and in the smallest-density case for even over $10^6 t_0$.

The gas-liquid critical point of our system was evaluated via grand-canonical Monte Carlo (GCMC) simulations and using the histogram reweighting technique~\cite{Wilding_JPCM_1997}. We consider a cubic box with side $L=11$ and looked for values of temperature ($T$) and chemical potential ($\mu$) at which the system exhibited large fluctuations in both the number of particles ($N$) and the total internal energy of the system ($U$).
At the state points where such fluctuations were found, we started ten independent MC runs -- each one extending over more than $10^6$ MC steps -- and evaluated the probability distribution of the order parameter $M\sim N + sU$, $s$ being the mixing parameter~\cite{Wilding_JPCM_1997}. The precise location of the critical point was then obtained by using the histogram reweighting technique~\cite{Wilding_JPCM_1997} and searching for those $(T,\mu,s)$ values at which the probability distribution of $M$ matched the one of the 3D Ising model~\cite{Wilding_JPCM_1997}. 

The theoretical description of the static properties of the 
model was carried out using the multi-density integral-equation theory developed in Ref.~\cite{yuraus}.
The pair distribution functions, the structure factors and the internal energy were calculated using numerical
solution of the corresponding multi-density Ornstein-Zernike equation supplemented by the APY closure relations. To avoid unnecessary repetition we refer the reader to the original 
publication~\cite{yuraus}, where the theory is described in details. 
In contrast to Ref.~\cite{yuraus} where the reference model has always a HS contribution, 
here we consider both a HS and a soft sphere core; the latter is the same used in MD simulations.
In the following, theoretical results for the soft sphere case are reported only if we observed significative differences with respect to the HS case. 

\section{Results}\label{sec:results}

In an effort to obtain a better overview over the system behavior we have first located the critical point of the gas-liquid transition via GCMC simulations, using the histogram reweighting technique~\cite{Wilding_JPCM_1997}; this point was found to be located at $(T_c^*,\mu_c^*, \rho_c^*)=(0.122,-0.477, 0.266)$. In matching the distribution of the order parameter $M$ to the binodal curve of the Ising model, the best fit was achieved for $s\approx 0$.

MD simulations were carried out on a grid of state points, defined by six densities ($\rho^* = 0.1, 0.25, 0.4, 0.55, 0.7$ and 0.8) and six temperature values ($T^*=0.10, 0.11, 0.12, 0.13, 0.15$ and 0.30); $\rho^* = 0.8$ was the highest density value for which we could equilibrate the system in the liquid phase.
Temperatures were chosen to span from super-critical to sub-critical values in such a way that the isotherms better sample the region around the critical point.

\subsection{Static structure and thermodynamics}
\label{sec:static}

To describe the static structure of the system we have calculated the pair distribution function, $g(r)$, defined by~\cite{hansenmcdonald}

\begin{equation}
\rho g(r)= \frac{2}{N} \left \langle 
\sum_{i=1}^N \sum_{j=i+1}^N \delta(\vec{r} - \vec{r}_j + \vec{r}_i) \right \rangle,
\end{equation}
and the static structure factor~\cite{hansenmcdonald}
\begin{equation}
\label{eqn:sk}
S(k) = \frac{1}{N} \left \langle
\left | \sum_{n=1}^N e^{i \vec{k}\cdot \vec{r}_n} \right |^2   \right \rangle,
\end{equation}
where $\vec{r}$ is the distance between two IPC centers, $\vec{r}_i$ with $i = 1, \cdots, N$ represents the position of particle $i$ and $k$ is the wavevector. $S(k)$ and $g(r)$ were calculated directly during the simulation, i.e., ``on the fly''.
 
While in the simulations $g(r)$ and $S(k)$ are obtained according to the aforementioned definitions, within the APY theory $g(r)$ is obtained from the Fourier transform of $S(k)$.

\begin{figure*} 
\begin{center}
\includegraphics[width = 1.0\textwidth]{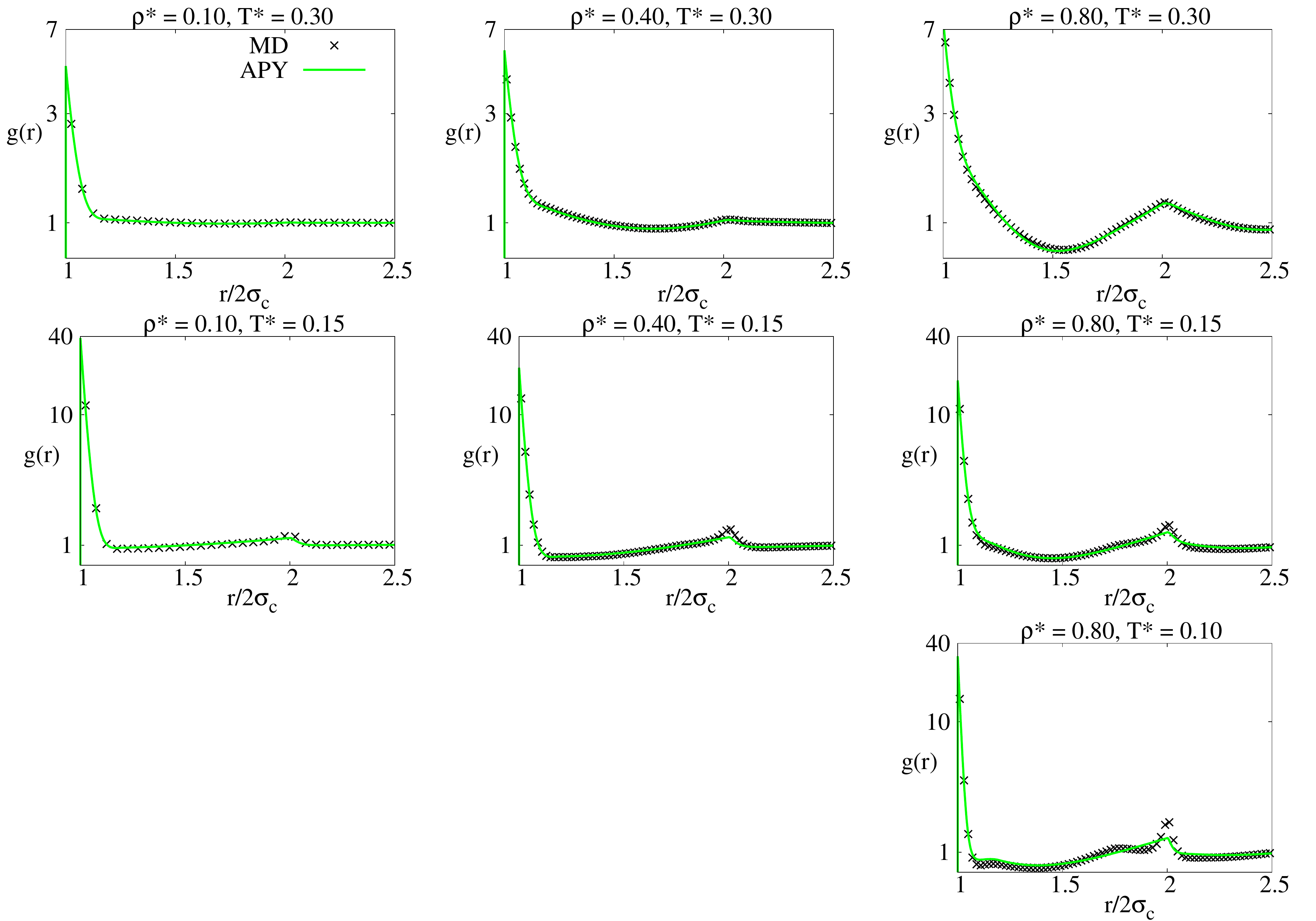}
\caption{
Pair distribution functions $g(r)$ of the system at different state points (as labeled).
Solid green lines correspond to APY results, crosses refer to MD data.
In the missing panels the system is phase separating.
\label{fig:g}}
\end{center}
\end{figure*}

In Figure~\ref{fig:g} data for $g(r)$ at the selected state points are reported. Each panel refers to a different thermodynamic state: the left column corresponds to low density cases, i.e. $\rho^*=0.10$, the central column displays data at an intermediate density, i.e. $\rho^*=0.40$, and the right column shows results for the highest density state points, i.e. $\rho^*=0.80$; temperature decreases from the top to the bottom, dropping from $T^*=0.30$ to $T^*=0.10$. Since we report here only equilibrium data, two low-temperature panels are missing in Figure~\ref{fig:g} because they correspond to state points located within the binodal.

Throughout, a good, sometimes even excellent, agreement between theoretical and simulation results can be observed; this holds in particular for distances covering the first neighbor shell, i.e. for $r\le 2 \sigma_{\rm c} + \delta$.
Since the APY approach -- as all integral equation approaches -- allows to calculate the thermodynamic properties by integrating $g(r)$ over the interaction range of the potential, a reliable evaluation of the pair distribution function within the first neighbor shell is indispensable for an accurate calculation of these properties.
In addition, a reliable prediction of the contact value of the pair distribution function, namely $g(2\sigma_{\rm c})$, is responsible for a reliable estimate of the pressure in systems with hard-core interactions.
At low densities and high temperatures (upper-left panel) the numerical $g(r)$ reproduces the characteristic features of a hard sphere system, characterized by a pronounced peak at contact, i.e., for $r = 2\sigma_{\rm c}$, and a fast, structureless decay to 1 on increasing $r$. Upon lowering the temperature and/or increasing the density, a second nearest neighbors peak gradually emerges at $r = 4 \sigma_{\rm c}$.
The overimposed theoretical data show that APY predictions are in good agreement with the simulation results not only at contact, but also around the second peak of $g(r)$.
Only for the high-density, low-temperature state point (bottom-right panel) small discrepancies can be observed: the theoretical data are not able to reproduce the complex shape of the peak at $r \simeq 4 \sigma_{\rm c}$.
In subsection~\ref{sec:dynamics} we will provide evidence that at this state point the system dynamically slows down.

\begin{figure*} 
\begin{center}
\includegraphics[width = 1.0\textwidth]{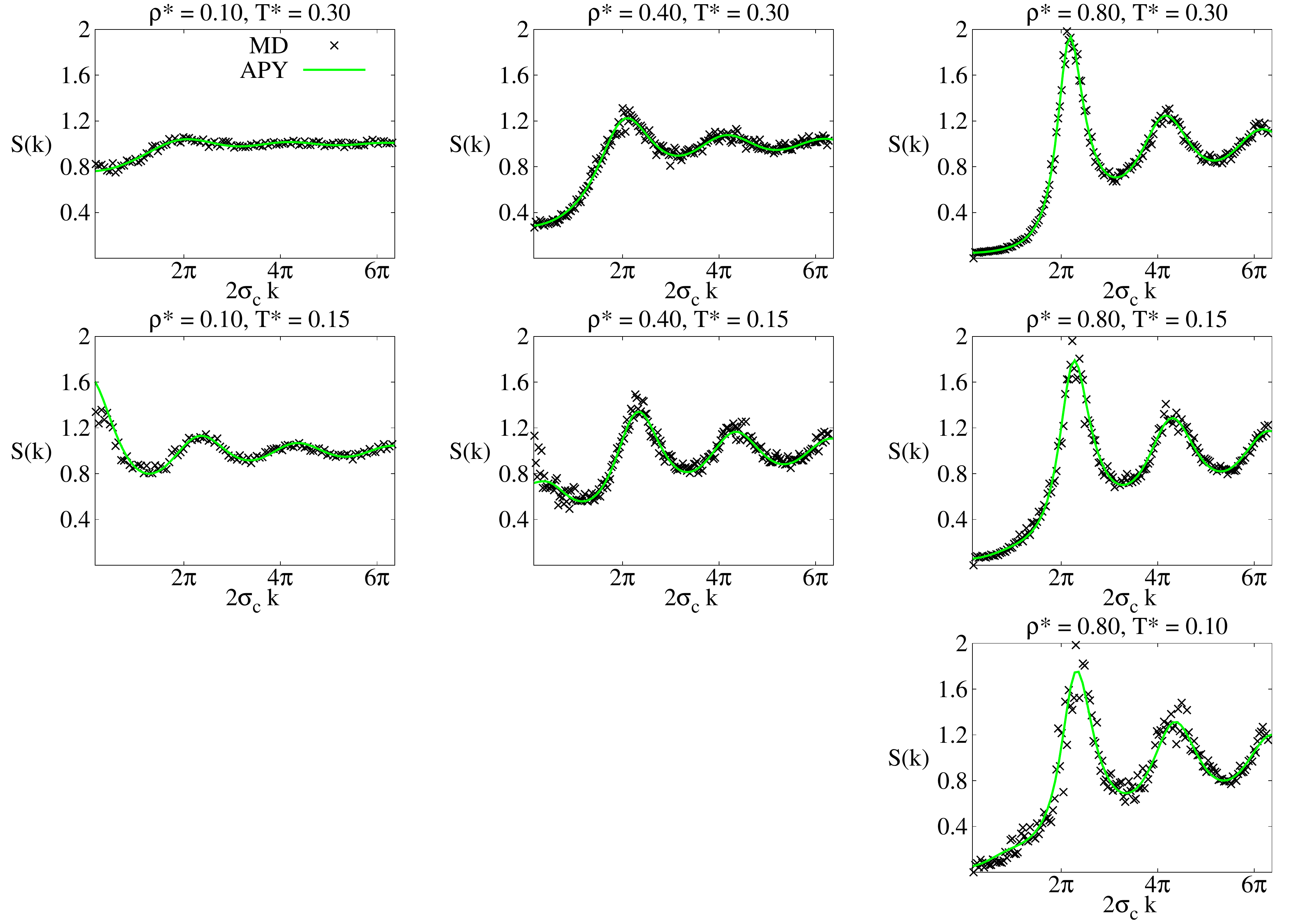}
\caption{
Static structure factor $S(k)$ of the system at different state points (as labeled).
Solid green lines correspond to APY results, crosses refer to MD data.
In the missing panels the system is phase separating. 
\label{fig:sk}}
\end{center}
\end{figure*}

In Figure~\ref{fig:sk} we show the corresponding $S(k)$, using the same arrangement of panels as in Figure~\ref{fig:g}.  Similar to the pair distribution function the agreement between theory and simulations is extremely good, even though data are slightly noisy due to substantial differences in the statistics.
While at high temperatures and/or high densities the theoretical and numerical data fall essentially on the top of each other, at low temperatures and low/intermediate densities slight discrepancies can be observed at $k = 0$. The value of the static structure factor at $k = 0$ is of particular relevance since $S(0)/\rho k_B T$ is the isothermal compressibility. 
An increase of $S(0)$ indicates the onset of a liquid-gas phase separation.
In our case at $T^*=0.15$, i. e. at a temperature slightly above the gas-liquid transition temperature, we observe the onset of an increase in $S(0)$ in the simulation data; the theoretical approach is able to reproduce this increase of the compressibility (with slight discrepancies).
In contrast, far from the gas-liquid phase separation region, theoretical predictions are very accurate, even at very high density.

\begin{figure} 
\begin{center}
\includegraphics[width = 1.0\textwidth]{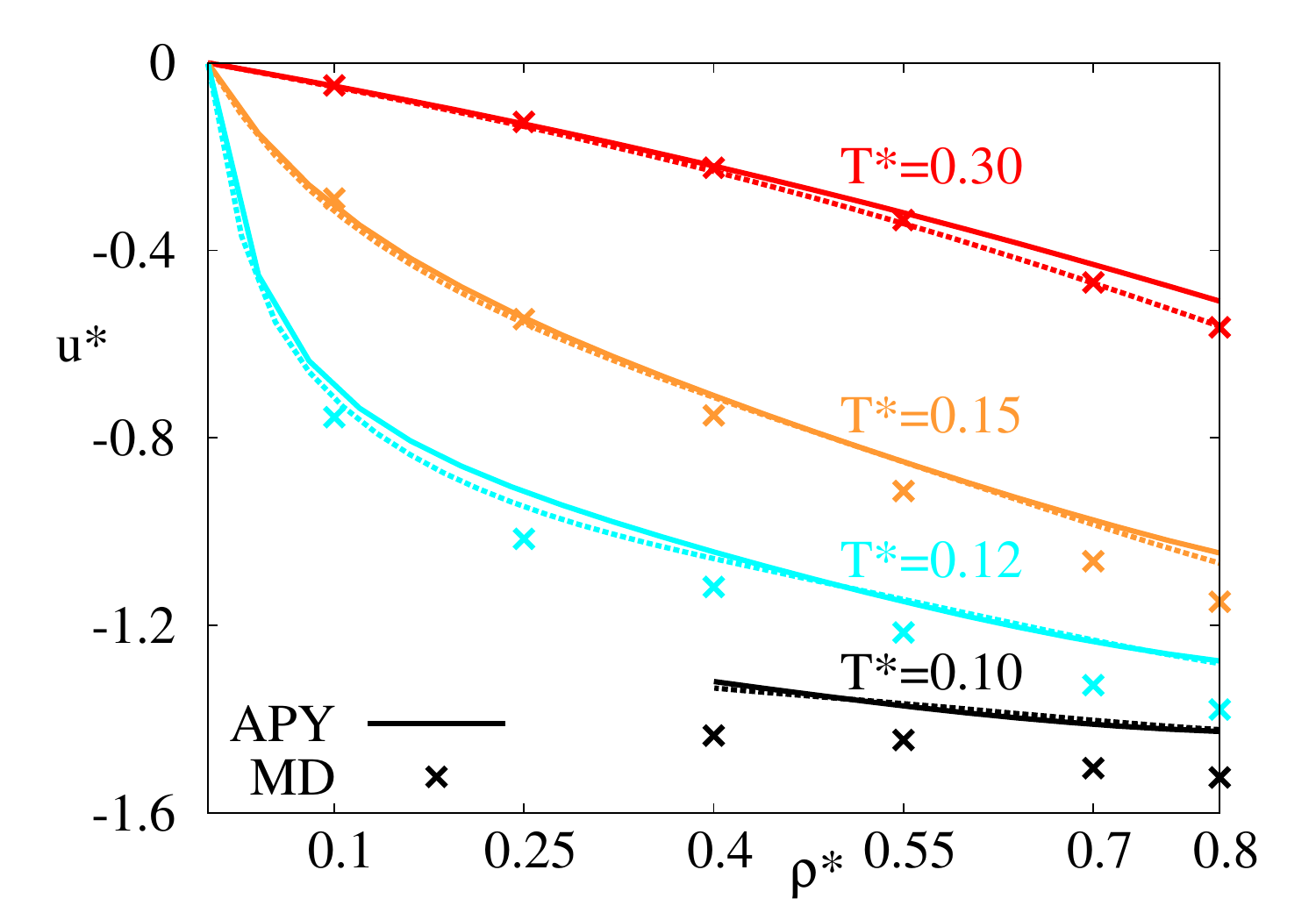}
\caption{
Potential energy per particle as a function of the density of the system along several isotherms (as labeled).
Crosses are MD results, solid lines refer to APY data for the system with the HS core, and dashed lines correspond to APY data for the system with the soft sphere core.
\label{fig:uxc}}
\end{center}
\end{figure}

Finally, in Figure~\ref{fig:uxc} we display the internal energy per particle versus the density of the system along four isotherms; data along the lowest temperature isotherm are incomplete in a density range where the system undergoes phase separation.
At the state points where APY converges, the agreement between the theoretical and numerical data is good for super-critical temperatures, while it is less convincing for sub-critical temperatures. We also report APY data for the soft sphere core. A small difference can be observed between the two reference models: the soft sphere provides a slightly more accurate agreement only at high temperatures, while at low temperatures the two theoretical sets of data reproduce simulation results with the same accuracy.

\subsection{Dynamics}\label{sec:dynamics}

We study the dynamics of the system exclusively via simulations by
calculating the autocorrelation functions of two particular
observables as well as the diffusion coefficient at several state
points.

To be more specific, we have calculated the autocorrelation function of
the particle velocities $\vec{v}(t)$, defined as
\begin{equation}
C_{\vec{v}}(t) = \frac{ \left \langle \vec{v}(t)\cdot \vec{v}(0) \right
  \rangle }{ \left \langle \vec{v}(0) \cdot \vec{v}(0) \right
  \rangle},
\end{equation}
as well as the autocorrelation function of the particle orientation
\begin{equation}
C_{\vec{n}}(t) = \left \langle \vec{n}(t)\cdot \vec{n}(0) \right \rangle,
\end{equation}
with $\vec{n}(t)$ being a unit vector pointing from one patch to
the other.
Both functions are single particle correlation functions; thus the
angular brackets denote a time average along the simulation as
well as an average over all particles.

\begin{figure*} 
\begin{center}
\includegraphics[width = 1.0\textwidth]{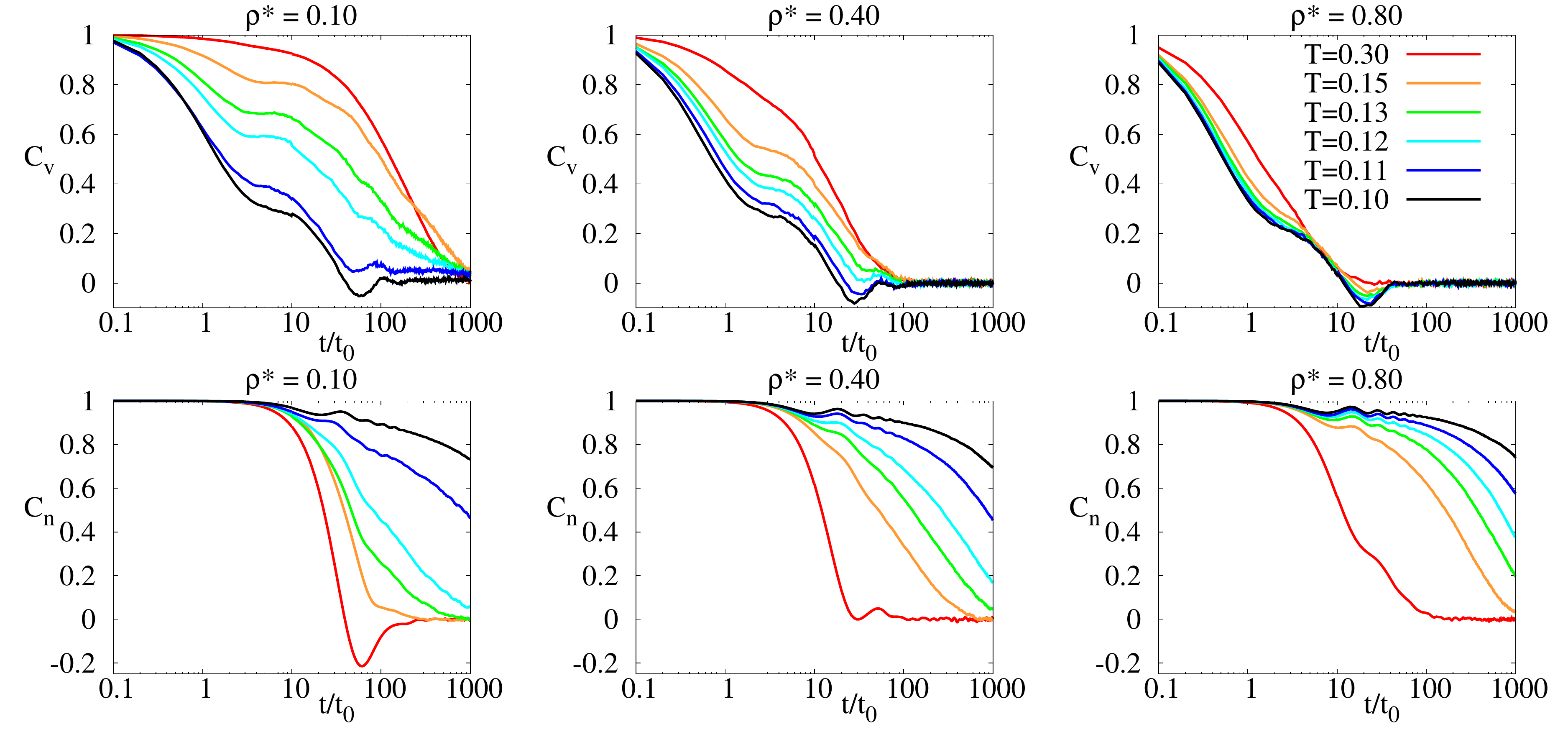}
\caption{
MD results for the velocity (top panel) and orientation (bottom panel) autocorrelation function at several temperatures (as labeled).
\label{fig:ac}}
\end{center}
\end{figure*}

Both correlation functions are shown in Figure \ref{fig:ac} for three
selected densities (considering a low, an intermediate and a high $\rho$-value) 
covering a range of temperatures from 0.3 down to 0.1. 

$C_{\vec{v}}(t)$ features the expected behavior:
(i) a monotonous decay at low/intermediate densities and high temperatures, and
(ii) a characteristic minimum, signaling a rebound mechanism, as the density increases and/or the temperature decreases.

In contrast, $C_{\vec{n}}(t)$ shows a rather complex, strongly
state-dependent behavior. At low temperatures and for all densities
this correlation function is slowly decaying, indicating a strong
persistence in the orientational motion of the particles. As the
temperature increases the decay of the correlation function is still
monotonous, indicating that the particles prefer to maintain their
original orientation; $C_{\vec{n}}(t)$ drops more rapidly with time as
the density decreases. Only at rather high temperatures and low/intermediate 
densities, the correlation function shows a pronounced
minimum; this feature signals that the particles start to freely rotate in space. 

Another feature of $C_{\vec{n}}(t)$ deserves particular attention, even
though a thorough discussion of this phenomenon requires a much more
comprehensive and systematic investigation that we postpone to a
future contribution~\cite{silvano}. In particular for low temperature states the
orientational correlation function shows characteristic short-range
oscillations in time, whose frequency increases with density. This
feature stems from fast oscillations of the particle orientation
around a preferred spatial axis and is undoubtedly a finger-print of
the strong coupling between the particles which occurs preferentially
at low temperatures (where particles are tightly bonded) and/or high
densities (where the average distance between particles is rather
small).  Also the fact that the frequency of the spatial oscillations
decreases with increasing density can be understood via a stronger
spatial coupling of the particles as $\rho$ grows.

Finally, we determined the diffusion coefficient, $D$, at all the
state points investigated. This quantity is calculated via the mean square
displacement of the particles, using the well-known relation
\begin{equation}
\left \langle \left [ \vec{r}(t)-\vec{r}(0) \right ]^2 \right \rangle \sim 6Dt
\end{equation}
at sufficiently large $t$ values, when the system is in the diffusive
regime.

\begin{figure*} 
\begin{center}
\includegraphics[width = 1.0\textwidth]{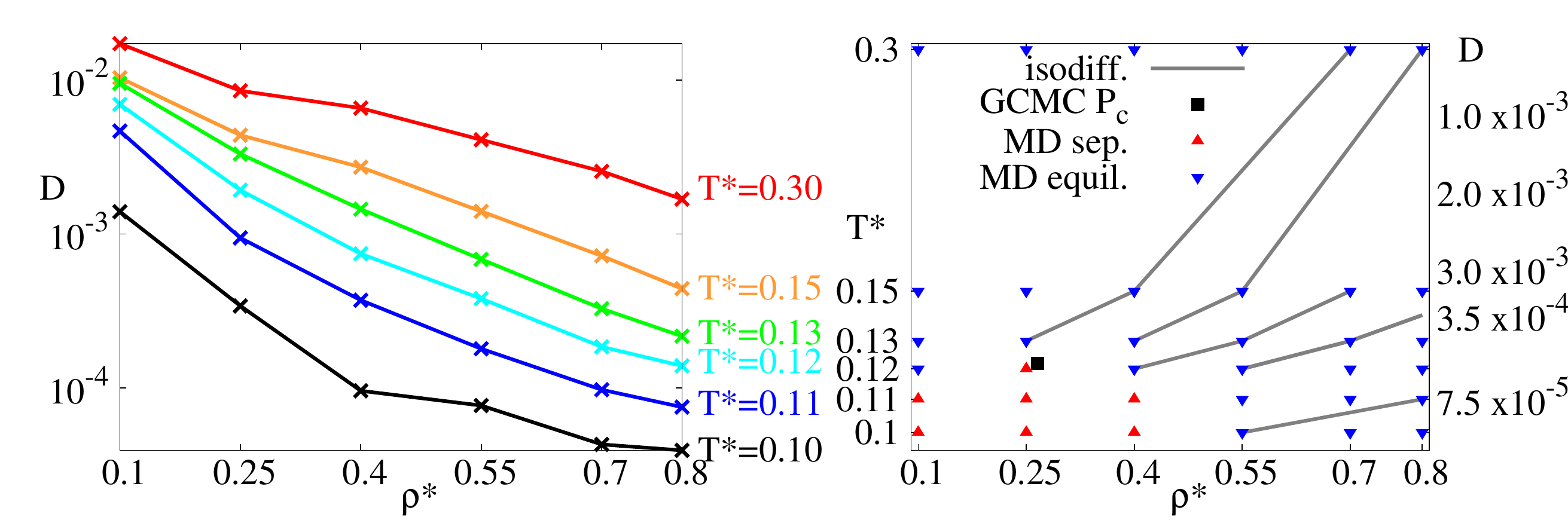}
\caption{
Left: MD results for the diffusion coefficient along several isotherms (as labeled). 
Right: phase diagram of the system. The points highlighted in the temperature-density plane correspond to the numerically investigated state points: a red \fulltriangle\ denotes a phase separating system, a blue \fulltriangledown\ corresponds to a homogeneous fluid, the black \fullsquare\ is the critical point.  Along the gray curves the value of the diffusion coefficient $D$ is constant and decreases from $10^{-3}$ to $7.5\cdot10^{-5}$ in units of $(2\sigma_{\rm c})^2/t_0$.}
\label{fig:dc}
\end{center}
\end{figure*}

Results for $D$ are shown in the left panel of Figure~\ref{fig:dc} as
a function of $\rho$ along several isotherms. The diffusion
coefficient shows a monotonous decrease of $D$ on decreasing both $T$
and/or $\rho$. From these data, we were able to trace isodiffusivity
lines, selecting $D$ values in a range from $10^{-3}$ to
$7.5 \times 10^{-5}$ in units of $(2\sigma_{\rm c})^2/t_0$.  In the right panel of
Figure~\ref{fig:dc} we report the resulting lines in the temperature
vs. density plane.
In the figure, the state points are characterized by different symbols to specify if they correspond to an equilibrium fluid phase or if they are characterized by a large value of $S(0)$, indicating a phase-separated system.

It is worth to observe that many of the simulated state points are found to be in the part of the phase diagram where dynamic slowing down occurs, characterized by a diffusion constant less than $10^{-4}$.
We also observe that the simulation data suggest a rather broad gas-liquid phase separation region, extending, e.g., from $\rho^*=0.10$ to $\rho^*=0.40$ at $T^*=0.10$.
We note that MD simulations are consistent with the estimate of the critical point provided by GCMC simulations.

\section{Conclusions} \label{sec:conclusions}

In the present contribution we considered overall neutral inverse patchy colloids (IPCs) with a charged equatorial belt and two oppositely charged polar patches defined by a relatively small opening angle.
While the same system has been investigated under planar confinement in Ref.~\cite{bianchi:2d2013}, we study here its bulk behavior in the fluid phase by means of numerical and theoretical approaches.
On one side, we performed extensive Molecular Dynamics (MD) simulations over a wide range of temperatures and densities, including extremely low temperatures, where a gas-liquid phase separation takes place, as well as high densities, where a dynamical slowing down of the system occurs. 
On the other side, we applied an integral equation approach developed to describe heterogeneously charged systems with two or more patches~\cite{yuraus}.
We found that this theory is able to predict the static and the thermodynamic properties of the systems with high accuracy within the investigated temperature-density range; this holds even at state points characterized by the slowing down of the dynamics.

Additionally, we numerically characterized the dynamics of the system by calculating the velocity and orientation autocorrelation function as well as the diffusion constant.
Finally, we traced a phase diagram where the gas-liquid phase separation region, the equilibrium fluid phase and the dynamically arrested state points are highlighted. We note that the observed bulk behavior, characterized by a wide region where the fluid phase is stable, is consistent with what observed in quasi two-dimensions.
Indeed, under planar confinement, the system was shown to form an extended network of bonded particles with no well-defined spatial order. In contrast, IPCs with relatively bigger patches, also studied under confinement in Ref.~\cite{bianchi:2d2013}, were shown to assemble into (spatially and orientationally) ordered planar arrays.
Thus, we expect to observe a completely different bulk phase diagram for the latter system~\cite{silvano}.

\section{Acknowledgments}
The authors gratefully acknowledge financial support by the Austrian
Science Fund (FWF) under Proj. Nos. V249-N27~(E.~B.),
P23910-N16~(S.~F.), and F41 -- SFB ViCoM (G. K).  The
work was supported within a bilateral ''Wissenschaftlich-Technisches
Abkommen'' project by the {\"O}AD under Proj. No. UA 04/2013.

\eject

\eject

\clearpage
\eject

\end{document}